\documentclass[]{emulateapj}

\usepackage{graphics,graphicx,xspace,natbib,amssymb}

\newcommand{\son}{SDSS\,J0912+1523\xspace}
\newcommand{\stt}{SDSS\,J2202-0033\xspace}
\newcommand{\oii}{[O\,{\sc ii}]\xspace}
\newcommand{\oiii}{[O\,{\sc iii}]\xspace}
\newcommand{\psb}{post-starburst galaxy\xspace}
\newcommand{\psbs}{post-starburst galaxies\xspace}

\newcommand{\arc}{\ensuremath{''}\xspace}

\newcommand{\msun}{\ensuremath{\rm{M}_\odot}\xspace}

\newcommand{\lprime}{\ensuremath{\rm{L}_{\rm{CO}}'}\xspace}
\newcommand{\kms}{\ensuremath{\rm{km\,s}^{-1}}\xspace}
\newcommand{\alphaco}{\ensuremath{\alpha_{\rm{CO}}}\xspace}

\newcommand{\Tdust}{\ensuremath{T_{\rm{dust}}}\xspace}

\newcommand{\Mgas}{\ensuremath{M_{\rm{gas}}}\xspace}

\shorttitle{Molecular Gas in Massive z$\sim$0.7 Post-Starburst Galaxies}
\shortauthors{Suess et al.}

\slugcomment{Accepted to ApJ Letters}

\begin{document}

\title{Massive quenched galaxies at \lowercase{z}~$\sim$~0.7 retain large molecular gas reservoirs}
\author{Katherine A. Suess\altaffilmark{1}, Rachel Bezanson\altaffilmark{2,3}, Justin S. Spilker\altaffilmark{4}, Mariska Kriek\altaffilmark{1}, Jenny E. Greene\altaffilmark{2}, Robert Feldmann\altaffilmark{5}, Qiana Hunt\altaffilmark{2}, Desika Narayanan\altaffilmark{6}} 

\altaffiltext{1}{Astronomy Department, University of California, Berkeley, CA 94720, USA}
\altaffiltext{2}{Department of Astrophysical Sciences, Princeton University, Princeton, NJ 08544, USA}
\altaffiltext{3}{Department of Physics and Astronomy and PITT PACC, University of Pittsburgh, Pittsburgh, PA 15260, USA}
\altaffiltext{4}{Steward Observatory, University of Arizona, Tucson, AZ 85721, USA}
\altaffiltext{5}{Institute for Computational Science, University of Zurich, Zurich CH-8057, Switzerland}
\altaffiltext{6}{Department of Astronomy, University of Florida, Gainesville, FL 32607, USA}
\email{suess@berkeley.edu}

\begin{abstract}
The physical mechanisms that quench star formation, turning blue star-forming galaxies into red quiescent galaxies, remain unclear. In this Letter, we investigate the role of gas supply in suppressing star formation by studying the molecular gas content of \psbs. Leveraging the wide area of the SDSS, we identify a sample of massive intermediate-redshift galaxies that have just ended their primary epoch of star formation. We present ALMA CO(2-1) observations of two of these \psbs at $z\sim0.7$ with $M_*\sim2\times10^{11}\,M_\odot$. Their molecular gas reservoirs of $(6.4\pm0.8)\times10^9\,M_\odot$ and $(34.0\pm1.6)\times10^9\,M_\odot$ are an order of magnitude larger than comparable-mass galaxies in the local universe. Our observations suggest that quenching does not require the total removal or depletion of molecular gas, as many quenching models suggest. However, further observations are required both to determine if these apparently quiescent objects host highly obscured star formation and to investigate the intrinsic variation in the molecular gas properties of \psbs.
\end{abstract}


\section{Introduction}
Galaxies at low and intermediate redshifts broadly fall into two categories: disky, blue star-forming galaxies and `quenched' red elliptical galaxies without ongoing star formation \citep[e.g.,][]{kauffmann03}. To create these observed galaxy populations, blue galaxies must somehow cease forming stars. Because stars form out of cold molecular gas, many quenching models attempt to explain the cessation of star formation through mechanisms that remove, deplete, or heat molecular gas reservoirs. For example, strong feedback from quasars could stop star formation by removing gas from the galaxy \citep[e.g.,][]{dimatteo05, hopkins06}, possibly induced by a major merger \citep[e.g.,][]{wellons15}. Massive molecular outflows, such as those observed locally by \citet{alatalo15} and \citet{leroy15}, could play a role in this process. Alternately, virial shocks could reduce the accretion of cool gas onto massive halos \citep[e.g.,][]{keres05, dekel06} and prevent new stars from forming. \citet{feldmann15} suggest that star formation suppression is initiated by the decline of cool gas accretion onto the galaxy once the parent dark matter halo switches from a fast collapsing mode to a slow accretion mode. 

To test these models, it is vital to investigate the molecular gas properties of galaxies just after they quench. \psbs-- whose spectra resemble A-type stars, indicating a burst of star formation that ended $\sim1$~Gyr ago \citep[e.g.,][]{leborgne06}-- represent the direct and unpolluted products of the quenching process. Low-redshift \psbs \citep[`E+A' or `K+A' galaxies, e.g.][]{dressler83,couch87,zabludoff96} have been shown to have larger molecular gas reservoirs than expected given their low star formation rates \citep[SFRs;][]{french15}. These low-mass galaxies are shutting off a small burst of late-time star formation, as opposed to quenching their major star-forming episode. Only by moving to higher stellar masses and redshifts can we study galaxies that have just finished their primary epoch of star formation \citep[e.g.,][]{whitaker12}; however, we must select galaxies at low enough redshifts that observations of low-J CO transitions are feasible.

In this Letter, we leverage the wide area of the Sloan Digital Sky Survey \citep[SDSS,][]{york00} to identify \psbs at $0.5< z<0.8$. We present ALMA CO(2-1) observations of two $z\sim0.7$ \psbs, which reveal large reservoirs of molecular gas.

We assume a cosmology of $\Omega_m=0.3,\ \Omega_\Lambda=0.7$, and $h=0.7$.

\section{Sample Selection}
To identify a sample of massive \psbs bright enough for follow-up study, we select galaxies from the SDSS DR12 \citep{dr12} spectroscopic catalog with $i<19$, $z\ge0.5$, and median spectral signal-to-noise ratio $>3.7$. We identify galaxies dominated by A-type stars following the method introduced in \citet{kriek2010}, which selects \psbs by their strong Balmer breaks and their blue slopes redward of the break. The final selection was checked by eye to remove stars and brown dwarfs, and consists of 50 \psbs at $0.5~\le~z~\le 0.8$ with a median redshift of $z=0.6$. 

To derive the stellar masses of the \psb sample, we fit the SDSS spectra with the stellar population synthesis fitting code FAST \citep{kriek2009} using the \citet{bc03} stellar population library, a \citet{chabrier03} initial mass function (IMF), a delayed exponential star formation history, and the \citet{kcDust} dust attenuation law. The mass was aperture corrected using the difference between the photometric magnitude in the SDSS $g$, $r$, and $i$ filters and the same magnitudes as measured from the optical spectra; this correction increased the adopted stellar mass by a mean factor of 1.5. 

We measured SFRs using the aperture-corrected line flux of the \oii$\lambda$3727 doublet and the SFR conversion in \citet{kennicuttReview}, adjusted to a Chabrier IMF. To measure the \oii flux, we modeled the region around the doublet as a single Gaussian centered at the mean wavelength of \oii plus a straight-line continuum with a free slope and intercept. The width of the Gaussian line was held equal to the stellar velocity dispersion, given in the MPA-JHU catalogs \citep{aihara11}. In all cases, the velocity dispersion resulted in a single Gaussian wider than the separation between the two lines in the \oii doublet. Uncertainties in the SFR were bootstrapped from 1,000 realizations of the \oii line flux fit. We corrected the SFR for dust attenuation using the best-fit dust attenuation value from the stellar population fit (median A$_{\text{v}}$=0.8~mag). 

\begin{figure}
    \centering
    \includegraphics[width=.5\textwidth]{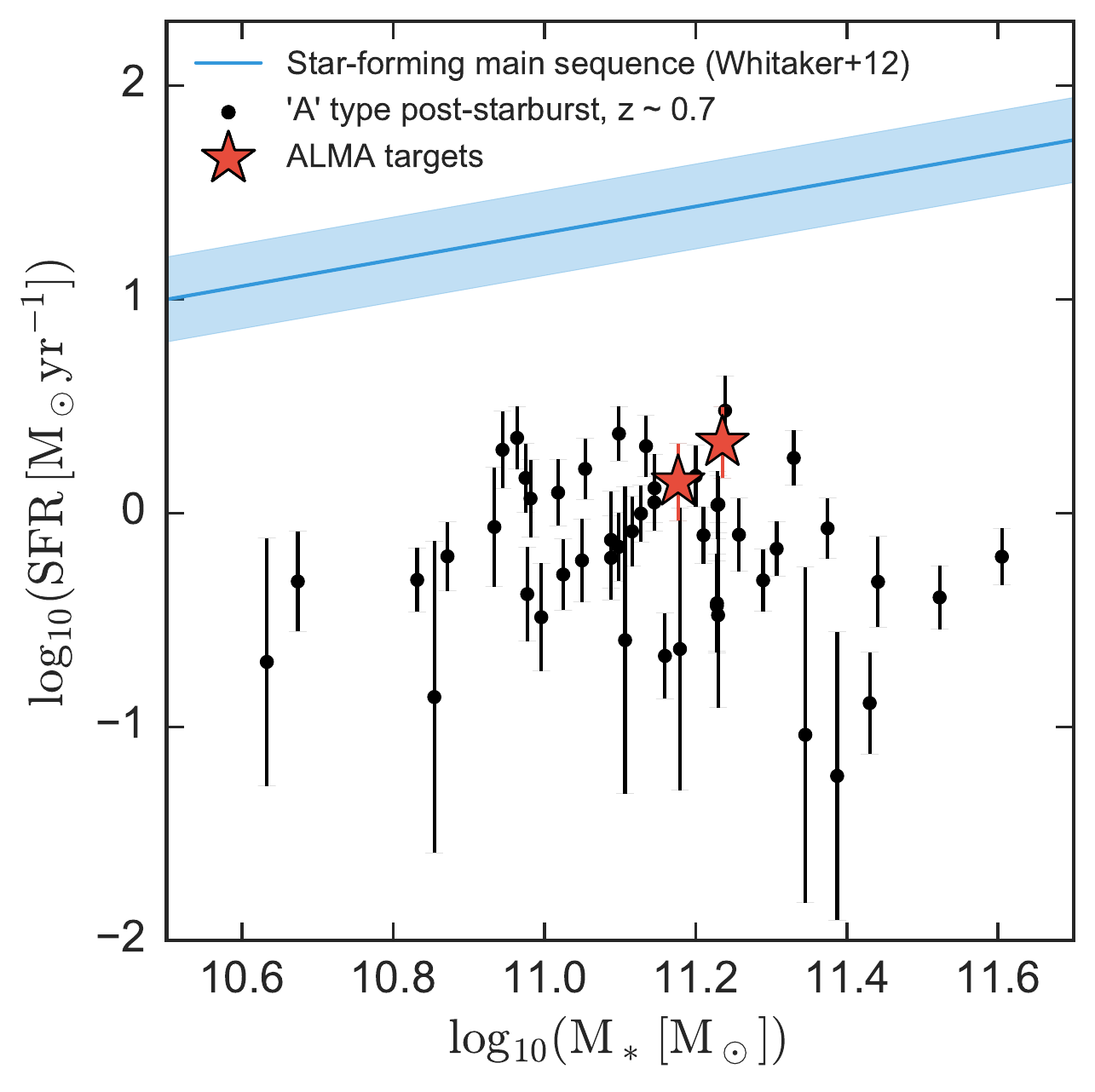}
    \caption{SFR, estimated using the dust-corrected \oii$\lambda$3727 luminosity, as a function of stellar mass for `A-type' \psbs selected from the SDSS. Red stars indicate the galaxies observed with ALMA. The blue shaded line indicates the star-forming main sequence at z=0.6 from \citet{whitakerMS}; all \psbs in the sample lie significantly below the main sequence.}
    \label{fig:SFR-M}
\end{figure}

Figure \ref{fig:SFR-M} shows the SFR as a function of stellar mass for the full \psb sample. All galaxies lie significantly below the star-forming main sequence at this redshift \citep{whitakerMS}. 

\section{ALMA Observations}

\begin{figure}
    \centering
    \includegraphics[width=.5\textwidth]{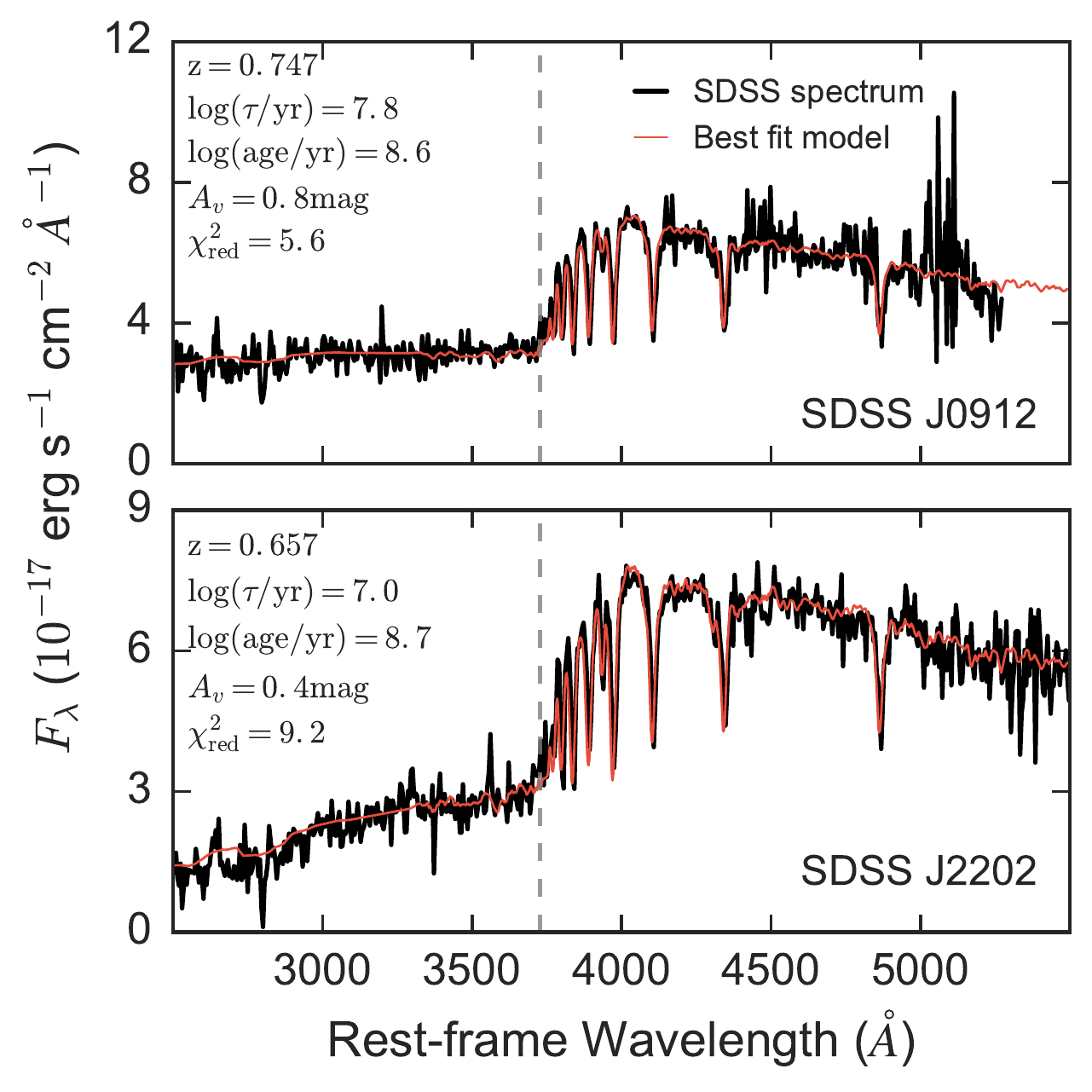}
    \caption{SDSS  discovery spectra of \son and \stt. Best-fit stellar population synthesis models are overplotted in red. The grey dashed line indicates the location of the \oii$\lambda$3727 doublet used to measure the SFR.}
    \label{fig:SDSSspectra}
\end{figure}

We selected two of the highest-redshift, brightest galaxies-- \son and \stt-- from the full sample for follow-up observations with ALMA. The SDSS discovery spectra of these two galaxies are shown in Figure \ref{fig:SDSSspectra}, and basic parameters are listed in Table \ref{t1}. The ALMA observations were carried out in program 2016.1.01126.S (PI: R.~Bezanson) in January and March 2017 using the ALMA Band~4 receivers \citep{asayama14}. Observations were made in 80\,min blocks, with two blocks dedicated to each target. The total on-source integration time for each target was $\sim$100\,min. The data were reduced using the standard ALMA pipeline, which produced good results for both \son observing blocks and one \stt block. In the second observation of \stt, many antennas were shadowed by other antennas during observations of the bandpass calibrator due to the compact array configuration. This caused the automatic pipeline to flag the shadowed antennas for the remainder of the observing block. We instead derived the frequency-dependent response of the affected antennas using the repeated observations of the complex gain calibrator carried out during the remainder of the track. While the gain calibrator is not as bright as the bandpass calibrator, we verified that the bandpass solutions derived in this way are consistent with those using the bandpass calibrator for the antennas which were not shadowed.

\begin{deluxetable*}{ccccccc}
\tablecolumns{7}
\tablecaption{Post-starburst targets \label{t1}}
\tablehead{\colhead{ID} & \colhead{z} & \colhead{$M_*$ [$M_\odot$]} & \colhead{SFR [$M_\odot$/yr]} & \colhead{Aperture Correction} & \colhead{$L_{\rm{CO}}$  [K\,km\,s$^{-1}$\,pc$^2$]} & \colhead{$M_{\rm{gas}} [M_\odot]$}}
\startdata
\son & 0.747 & (1.7~$\pm$~0.3)$\times10^{11}$ & 2.1 $\pm$ 0.8 & 1.13$\pm$0.01 & 8.5 $\pm$ 0.4 & (34.0$\pm$1.6)$\times10^9$ \\
\stt & 0.657 & (1.5~$\pm$~0.2)$\times10^{11}$ & 1.4~$\pm$~0.6 & 1.11$\pm$0.01 & 1.6~$\pm$~0.2 & (6.4$\pm$0.8)$\times10^9$
\enddata
\tablecomments{Errors in $M_*$ are dominated by systematics, which we estimated by varying the model library, star formation history, and dust law in the stellar population fits.}
\end{deluxetable*}

The observations reach a spatial resolution of 2.2$\times$3.0\arc (16$\times$22$\,$kpc) and 1.7$\times$2.4\arc (12$\times$17$\,$kpc) for \stt and \son, respectively, in images inverted using natural weighting, which maximizes sensitivity at the expense of slightly lower spatial resolution. No 2\,mm continuum emission was detected in either source, with a 3$\sigma$ upper limit $<75$\,$\mu$Jy. 

CO(2--1) is significantly detected in both target galaxies. Inspection of the image cubes and a comparison of the maximum pixel values with the integrated flux density indicate that \son is marginally spatially resolved, with a deconvolved source FWHM of 1.9~$\pm$~0.3\arc (14~$\pm$~2$\,$kpc). \stt is not spatially resolved at the depth and resolution of our data. To extract spectra, we fit a point source (\stt) or a circular Gaussian (\son) to the visibility data using the \textit{uvmultifit} package \citep{martividal14}, averaging 24 (\stt) or 6 (\son) channels, yielding a velocity resolution of $\approx\,$200 and 50\,\kms, respectively. These spectra, as well as images integrated over the full linewidths of each target, are shown in Figure~\ref{fig:almaspectra}.

\section{Molecular Gas Masses}

\begin{figure}
\centering
\includegraphics[width=.5\textwidth]{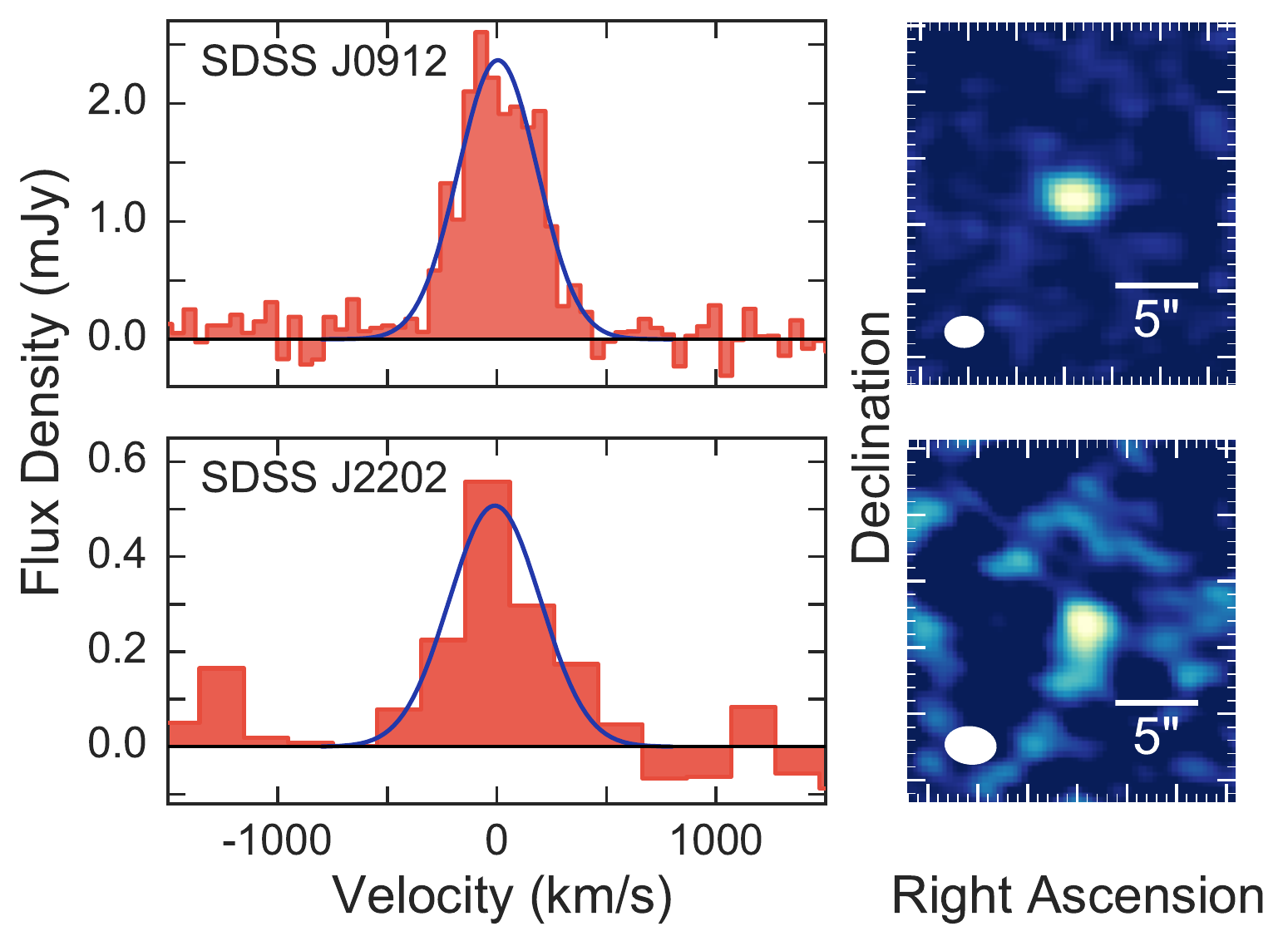}
\caption{Left: ALMA CO(2--1) spectra for each target. The blue line shows the best-fit Gaussian used to extract the total line flux. Right: integrated line images for each target. The white oval shows the ALMA beam.}
\label{fig:almaspectra}
\end{figure}

Both sources are significantly detected in CO(2--1) emission (see Figure~\ref{fig:almaspectra}). We fit simple Gaussian profiles to these spectra, which yield integrated line fluxes of $1.07\pm0.05$ and $0.27\pm0.03$\,Jy\,\kms for \son and \stt, respectively. At the redshifts of each source, these fluxes correspond to line luminosities $\lprime=8.5\pm0.4$ and $1.6\pm0.2 \times 10^9$\,K\,\kms\,pc$^2$, respectively.

The CO line luminosity can be converted to a molecular gas mass using the conversion factor \alphaco. In high metallicity objects, \alphaco typically varies from $\approx0.8$\,\msun/(K\,\kms\,pc$^2$) in highly star-forming objects (typically gas-rich merging systems) to $\approx4$\,\msun/(K\,\kms\,pc$^2$) in the Milky Way and other normal star-forming galaxies, principally due to the escape probability in the optically-thick CO line (see \citealt{bolatto13} for a recent review; hereafter we suppress the units of \alphaco). 

We estimate the value of \alphaco using the numerical models of \citet{narayanan12}. These authors combined hydrodynamical modeling of a variety of isolated and merging galaxies with molecular line radiative transfer to develop a fitting formula for \alphaco as a function of the CO surface brightness. \stt is not spatially resolved at the depth and resolution of our data, which yields a lower limit on the CO surface brightness and an upper limit on the conversion factor of $\alphaco\lesssim7.0$. Our data spatially resolve \son; combined with its CO luminosity, this implies $\alphaco\sim3.5$. For ease of comparison with the literature, we therefore adopt \alphaco=~4.0 for both of our targets. 

Additionally, \alphaco is valid and applies only to the CO(1--0) transition, so observations of higher CO lines require a correction factor to the ground state. Based on observations of a wide variety of galaxies at many redshifts, the CO(2--1)/CO(1--0) line ratio likely falls in the range 0.7--1.0 in temperature units, where 1.0 indicates thermalized emission \citep[e.g.,][]{combes07,dannerbauer09,young11}. We assume thermalized emission in our analysis. This is a conservative assumption, yielding a minimum CO(1--0) line luminosity; the true values may be higher by $\approx\,$30\%.

Given these assumptions, we find molecular gas masses of $\Mgas=(34.0\pm1.6)\times10^9$\,\msun for \son and $\Mgas=(6.4\pm0.8)\times10^9$\,\msun for \stt. These uncertainties reflect only the statistical uncertainties on the measured CO luminosities; we stress that there are also factor of two systematic uncertainties associated with the CO excitation and conversion factor.

\begin{figure}
    \centering
    \includegraphics[width=.5\textwidth]{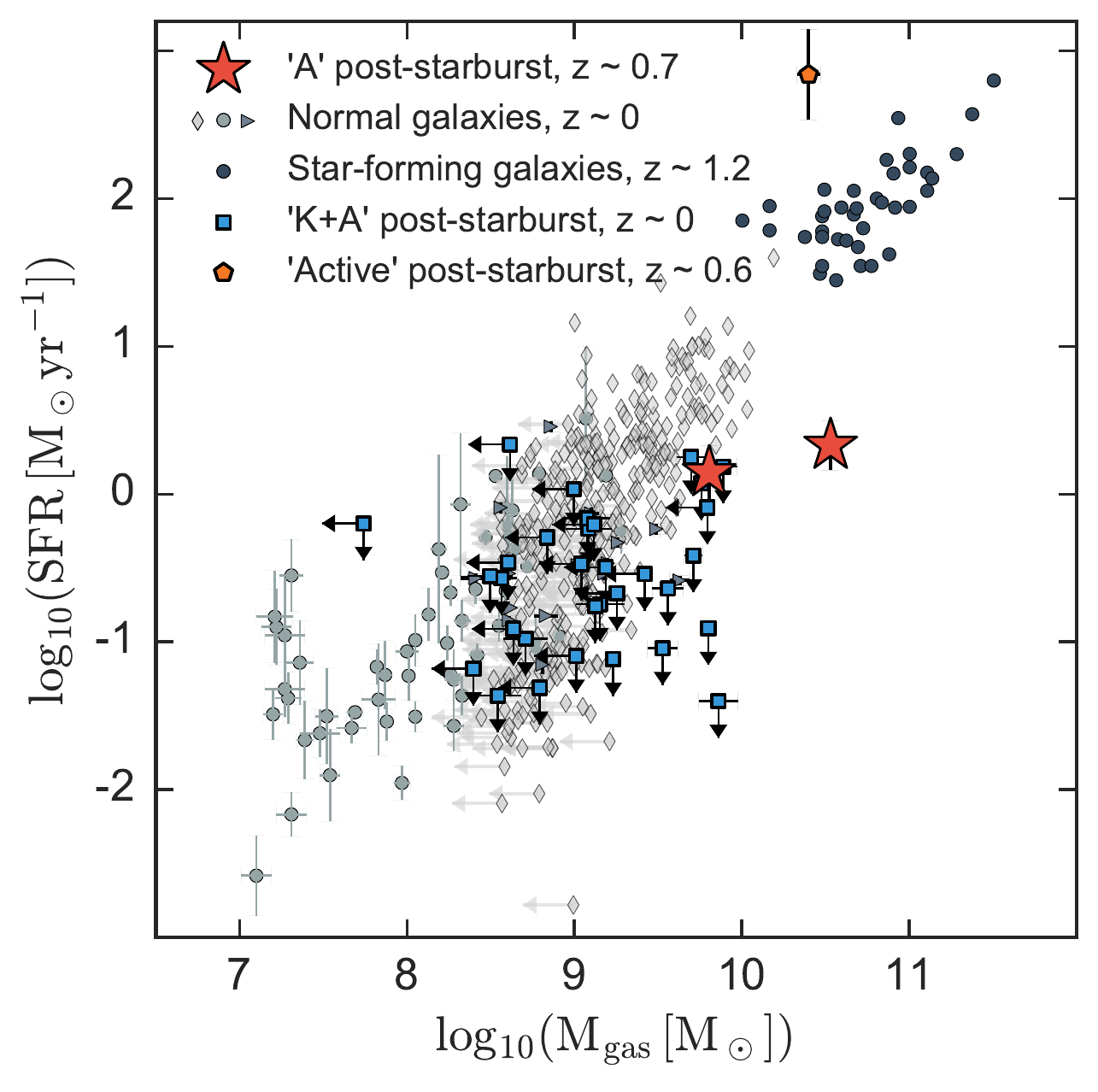}
    \caption{SFR as a function of molecular gas mass for \stt and \son (red stars) as well as comparison samples. Blue squares indicate the \citet{french15} low-redshift `K+A' \psbs; all SFRs are upper limits due to possible LINER emission. Dark grey points are star-forming galaxies from PHIBSS at $z\sim1.2$. Grey diamonds are normal galaxies from COLDGASS, grey circles are from ATLAS-3D, and grey triangles are from MASSIVE. The orange pentagon is an `active' \psb identified in \citet{sell14}, with a gas mass measured by \citet{geach13}. All have been normalized to \alphaco$=4$. \son and \stt are clearly offset from star-forming galaxies at both $z\sim0$ and $z\sim1.2$.}
    \label{fig:SFR-Mgas}
\end{figure}

In Figure~\ref{fig:SFR-Mgas}, we show SFR versus molecular gas mass for \son and \stt. It is clear that our observations, especially \son, are significantly offset from measurements of star-forming galaxies at both low and high redshifts. Adopting the CO size, the gas mass surface density of \son is a factor of 11 higher than expected from its SFR surface density \citep{kennicutt98}. \son has a gas mass a factor of two larger than expectations from the empirical scaling relation in \citet{genzel15}; \stt has a gas mass three times lower than expected from the same scaling. 

In Figure~\ref{fig:SFR-Mgas} we show for comparison the low-redshift `K+A' \psbs in \citet{french15}, an `active' \psb in the \citet{sell14} sample observed in CO by \citet{geach13}, and star-forming and quiescent galaxies at $z\sim 0$ (COLD GASS, \citealt{saintonge11}; ATLAS-3D, \citealt{young11,cappellari11,cappellari13,davis13,davis14}; and MASSIVE, \citealt{davis16}) as well as $z\sim 1.2$ (PHIBSS, \citealt{tacconi13}). Molecular gas fractions ($f_{\rm{gas}} \equiv M_{\rm{gas}} / M_*$) vary significantly both amongst the three post-starburst samples and within in each sample: the K+A post-starbursts have $f_{\rm{gas}} \sim 1-40\%$, the active post-starburst has $f_{\rm{gas}} \sim 19\%$, and \son and \stt have respective $f_{\rm{gas}}\sim$  20\% and 4\%.

\section{Implications for Galaxy Quenching}
Our results imply that star formation can be suppressed in galaxies that still have significant reservoirs of molecular gas. Models that require the complete depletion or ejection of molecular gas to quench star formation are inconsistent with the large gas masses in \son and \stt. Models that explain star formation suppression without removing the gas-- such as morphological quenching \citep{martig09}, which suggests that a buildup of stars into a bulge can stabilize some cold molecular gas against collapse-- are more consistent with our observations. However, it is not clear that current quenching models can fully explain the observed elevated molecular gas masses, especially for \son. 

Alternatively, the star formation in \son and \stt may only be temporarily suppressed. Simulations of massive z$\sim$2 galaxies predict a large amount of variability in ISM properties and star formation activity on short ($\sim$tens of Myr) timescales \citep{feldmann17}. These simulations contain a small number of galaxies with molecular gas fractions and SFR/$M_*$ similar to \stt. However, they are not able to reproduce a counterpart of the more extreme case of \son.
  
The factor of five difference in gas mass between \son and \stt indicates that, unsurprisingly, quenching may proceed differently from galaxy to galaxy. A larger galaxy sample is required to investigate intrinsic variations in the gas properties of \psbs and determine whether \son is an intriguing outlier or representative of recently quenched galaxies.

\section{Comparison to other studies}
Our results are similar to \citet{french15}, who found high molecular gas masses given the SFR of `K+A' \psbs at $0.01<z<0.12$. These authors interpret their observations as evidence for suppressed star-formation efficiency, a low \alphaco, or a bottom-heavy IMF in these galaxies.

\begin{figure*}
    \centering
    \includegraphics[width=\textwidth]{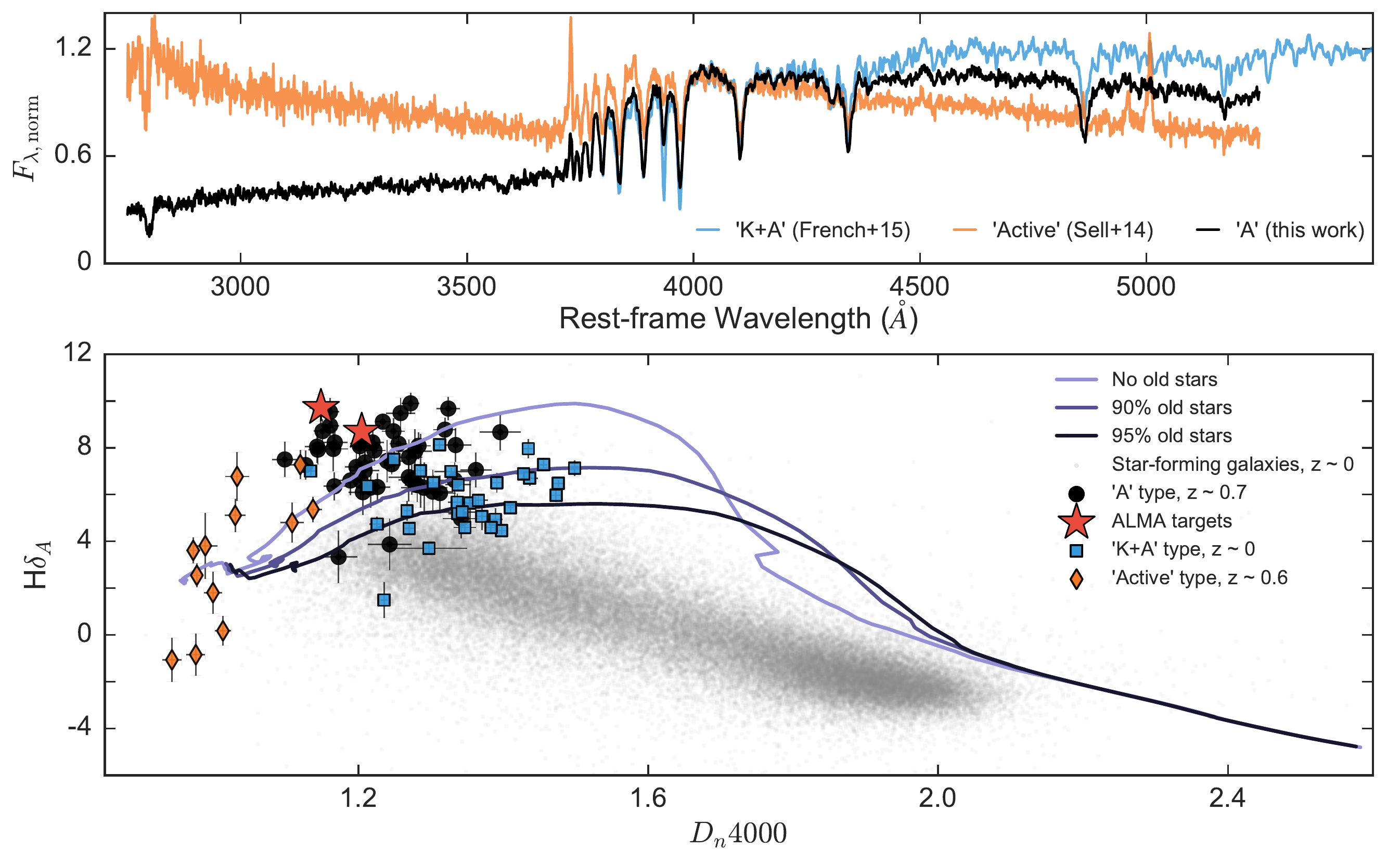}
    \caption{Top: median stacked spectra of `K+A' \citep[blue]{french15}, `active' \citep[orange]{sell14}, and `A' type (black) \psbs.  Bottom: H$\delta_A$ vs $D_n4000$ for the same three post-starburst samples. ALMA targets are shown with red stars; SDSS galaxies are shown in grey. Purple lines indicate \citet{bc03} models where 0, 90, and 95\% of stars were formed in a previous burst. As time increases, the model tracks evolve left to right.}
    \label{fig:Hd-D4000}
\end{figure*}

There are, however, significant differences between the \citet[hereafter `K+A sample']{french15} \psbs, the \psbs in this study, and the \psbs in other studies such as \citet[hereafter `active sample']{sell14}. The K+A sample includes $z<0.12$, $M_*\sim10^{10.6}\,M_\odot$ galaxies with strong Balmer lines and low ongoing star formation. The active sample includes $z\sim0.6$, $M_*=10^{10.5}$~-~$10^{11.5}\,M_\odot$ galaxies with strong Balmer absorption and weak nebular emission. In this study, we select $z\sim0.6$, $M_*\sim10^{11.1}\,M_\odot$ galaxies with strong Balmer breaks and blue slopes redward of the break. Despite all selections emphasizing `post-burst' criteria, the median stacked optical spectra of the three samples (top panel of Figure~\ref{fig:Hd-D4000}) are clearly distinct. The active sample shows prominent UV continuum as well as \oii and \oiii emission lines; inferred SFRs are $\sim 300~M_\odot \ \text{yr}^{-1}$. The K+A sample does not show UV continuum or emission lines, but has excess emission at rest-frame $\lambda~\ga~4500~\AA$ indicating a larger fractional contribution from old stars than the `A-type' post-starburst selection in this Letter. 

To further illustrate the differences between the post-starburst samples, we show the H$\delta_A$ index as a function of $D_n4000$ for all three samples as well as SDSS DR12 galaxies with $0.05\le~z\le 0.07$ and $M_\odot\ge10^{10}M_\odot$ (bottom panel of Figure~\ref{fig:Hd-D4000}). The H$\delta_A$ index traces recent star formation, and $D_n4000$ is sensitive to the age of the stellar population. For all three post-starburst samples we measured H$\delta_A$ using EZ\_Ages \citep{graves08}, adopting the Princeton 1D VDISP measurement for the velocity dispersion correction; we measured $D_n4000$ from the SDSS spectra \citep{balogh99}. Purple lines of increasing shade indicate the time evolution of \citet{bc03} single/double-burst models with an increasing contribution from old stars. The first model is a simple stellar population; the second is a model where 90\% of the stars were formed in a starburst 4~Gyr before the current star formation epoch; the third is a model where 95\% of the stars were formed in a previous starburst. 

The active sample lies at low $D_n4000$ and has a range of H$\delta_A$ values likely corresponding to emission partially filling in the $H\delta$ absorption feature. Given their high inferred SFR, these galaxies are still `active' and have not yet fully quenched their star formation; it is possible that after they quench, these galaxies will resemble A-type \psbs. The K+A post-starbursts are consistent with models where only 5-10\% of the galaxy's mass was formed in the most recent burst. This indicates that these galaxies formed the majority of their stars at higher redshift, and have just quenched a small amount of late-time star formation. Their spectra show the signatures of recently formed A-type stars as well as older K-type stars from previous star formation episodes. The A-type post-starbursts selected in this Letter are consistent with a simple stellar population model; i.e., they likely just finished their primary epoch of star formation and thus look like A-type stars. 

Given these differences, it is not clear that the same physical mechanisms should quench both the late-time star formation in the $z\sim0$ K+A galaxies and the primary epoch of star formation in the $z\sim0.7$ A-type \psbs. It is surprising that both the low-redshift K+A post-starbursts and intermediate-redshift A-type post-starbursts show high gas molecular gas masses for their SFR, indicating quenching in both redshift regimes can occur without exhausing or depleting molecular gas reservoirs. 

\section{Discussion}
In this Letter, we present ALMA CO(2-1) observations of two recently-quenched \psbs at $z\sim0.7$. These galaxies were selected from a unique sample of 50 A-type \psbs at $z\sim0.6$. Unlike low-redshift K+A \psbs, these massive galaxies have just concluded their major star-forming episode. Additionally, this sample lies at low enough redshift that follow-up observations of low-J CO transitions are feasible. The observed molecular gas masses in these galaxies are significantly higher than expected given their low ongoing SFR. This indicates that galaxies may quench their primary epoch of star formation without  completely depleting or exhausting their molecular gas. 

However, there are several alternative explanations for the high gas masses in \son and \stt. Short-term SFR variations, seen in some higher-redshift galaxy simulations, could account for the elevated gas mass of \stt. As in \citet{french15}, a lower \alphaco value could decrease the inferred gas mass and a bottom-heavy IMF could increase the inferred SFR, alleviating tension between the gas supply and low ongoing SFR. It is also possible that there is significant obscured star formation in \son or \stt.

We place a constraint on the dust-obscured star formation of each target using the non-detection of the 2mm ALMA continuum. We assume that the dust emission follows a standard modified blackbody function with $\Tdust=30$\,K \citep[typical of star-forming galaxies; e.g.,][]{magdis12}. The resulting upper limits on the FIR luminosity imply 3$\sigma$ upper limits on the obscured SFR\,$\lesssim50$\,\msun/yr. We additionally constrain the obscured star formation with WISE 12$\mu$m data \citep{mainzer11}, using the log average of \citet{dale02} templates with $1\le\alpha\le 2.5$ to convert the observed 12$\mu$m luminosity to a total IR luminosity, following \citet{whitakerMS}; this gives an upper limit of $\sim60\,\msun/yr$. 
Just $\sim6~M_\odot$/yr of obscured star formation-- well below the current constraints from either WISE or 2mm continuum-- would place \stt in line with the expected relation. In contrast, $\sim60~M_\odot$/yr of obscured star formation is required to account for the significant gas reservoir in \son. Both the radio and IR SFR upper limits are highly uncertain, and deeper rest-frame infrared measurements are required to place stronger constraints on any obscured star formation. 

Finally, we note that the sample size of this study is very small; measurements of the molecular gas in a larger sample of \psbs are necessary to determine if these galaxies are simply outliers. Thus, moving forward we need to obtain a larger sample of galaxies with gas measurements and tight constraints on the dust-obscured SFR. By studying the intrinsic variation in molecular gas properties of `A-type' \psbs, we may be able to constrain the primary mechanisms behind galaxy quenching.

\acknowledgements{
This paper makes use of the following ALMA data: ADS/JAO.ALMA~\#2016.1.00126.S.  ALMA is a partnership of ESO (representing its member states), NSF (USA) and NINS (Japan), together with NRC (Canada), NSC and ASIAA (Taiwan), and KASI (Republic of Korea), in cooperation with the Republic of Chile. The Joint ALMA Observatory is operated by ESO, AUI/NRAO and NAOJ. This material is based upon work supported by the National Science Foundation Graduate Research Fellowship Program under Grant No.~DGE~1106400.  KAS also acknowledges support from the University of California, Berkeley Chancellor's Fellowship. JEG is supported in part by NSF AST-1411642. RF acknowledges financial support from the Swiss National Science Foundation (grant 157591). DN was supported by NSF AST-1724864 and HST AR-13906.001.
}


\begin{thebibliography}{}
\expandafter\ifx\csname natexlab\endcsname\relax\def\natexlab#1{#1}\fi
\providecommand{\url}[1]{\href{#1}{#1}}

\bibitem[{{Aihara} {et~al.}(2011){Aihara}, {Allende Prieto}, {An}, {Anderson},
  {Aubourg}, {Balbinot}, {Beers}, {Berlind}, {Bickerton}, {Bizyaev}, {Blanton},
  {Bochanski}, {Bolton}, {Bovy}, {Brandt}, {Brinkmann}, {Brown}, {Brownstein},
  {Busca}, {Campbell}, {Carr}, {Chen}, {Chiappini}, {Comparat}, {Connolly},
  {Cortes}, {Croft}, {Cuesta}, {da Costa}, {Davenport}, {Dawson}, {Dhital},
  {Ealet}, {Ebelke}, {Edmondson}, {Eisenstein}, {Escoffier}, {Esposito},
  {Evans}, {Fan}, {Femen{\'{\i}}a Castell{\'a}}, {Font-Ribera}, {Frinchaboy},
  {Ge}, {Gillespie}, {Gilmore}, {Gonz{\'a}lez Hern{\'a}ndez}, {Gott}, {Gould},
  {Grebel}, {Gunn}, {Hamilton}, {Harding}, {Harris}, {Hawley}, {Hearty}, {Ho},
  {Hogg}, {Holtzman}, {Honscheid}, {Inada}, {Ivans}, {Jiang}, {Johnson},
  {Jordan}, {Jordan}, {Kazin}, {Kirkby}, {Klaene}, {Knapp}, {Kneib},
  {Kochanek}, {Koesterke}, {Kollmeier}, {Kron}, {Lampeitl}, {Lang}, {Le Goff},
  {Lee}, {Lin}, {Long}, {Loomis}, {Lucatello}, {Lundgren}, {Lupton}, {Ma},
  {MacDonald}, {Mahadevan}, {Maia}, {Makler}, {Malanushenko}, {Malanushenko},
  {Mandelbaum}, {Maraston}, {Margala}, {Masters}, {McBride}, {McGehee},
  {McGreer}, {M{\'e}nard}, {Miralda-Escud{\'e}}, {Morrison}, {Mullally},
  {Muna}, {Munn}, {Murayama}, {Myers}, {Naugle}, {Neto}, {Nguyen}, {Nichol},
  {O'Connell}, {Ogando}, {Olmstead}, {Oravetz}, {Padmanabhan},
  {Palanque-Delabrouille}, {Pan}, {Pandey}, {P{\^a}ris}, {Percival},
  {Petitjean}, {Pfaffenberger}, {Pforr}, {Phleps}, {Pichon}, {Pieri}, {Prada},
  {Price-Whelan}, {Raddick}, {Ramos}, {Reyl{\'e}}, {Rich}, {Richards}, {Rix},
  {Robin}, {Rocha-Pinto}, {Rockosi}, {Roe}, {Rollinde}, {Ross}, {Ross},
  {Rossetto}, {S{\'a}nchez}, {Sayres}, {Schlegel}, {Schlesinger}, {Schmidt},
  {Schneider}, {Sheldon}, {Shu}, {Simmerer}, {Simmons}, {Sivarani}, {Snedden},
  {Sobeck}, {Steinmetz}, {Strauss}, {Szalay}, {Tanaka}, {Thakar}, {Thomas},
  {Tinker}, {Tofflemire}, {Tojeiro}, {Tremonti}, {Vandenberg}, {Vargas
  Maga{\~n}a}, {Verde}, {Vogt}, {Wake}, {Wang}, {Weaver}, {Weinberg}, {White},
  {White}, {Yanny}, {Yasuda}, {Yeche}, \& {Zehavi}}]{aihara11}
{Aihara}, H., {Allende Prieto}, C., {An}, D., {et~al.} 2011, \apjs, 193, 29

\bibitem[{{Alam} {et~al.}(2015){Alam}, {Albareti}, {Allende Prieto}, {Anders},
  {Anderson}, {Anderton}, {Andrews}, {Armengaud}, {Aubourg}, {Bailey}, \&
  et~al.}]{dr12}
{Alam}, S., {Albareti}, F.~D., {Allende Prieto}, C., {et~al.} 2015, \apjs, 219,
  12

\bibitem[{{Alatalo}(2015)}]{alatalo15}
{Alatalo}, K. 2015, \apjl, 801, L17

\bibitem[{{Asayama} {et~al.}(2014){Asayama}, {Takahashi}, {Kubo}, {Ito},
  {Inata}, {Suzuki}, {Wada}, {Soga}, {Kamada}, {Karatsu}, {Fujii}, {Obuchi},
 {Kawashima}, {Iwashita}, \& {Uzawa}}]{asayama14}
{Asayama}, S., {Takahashi}, T., {Kubo}, K., {et~al.} 2014, \pasj, 66, 57

\bibitem[{{Balogh} {et~al.}(1999){Balogh}, {Morris}, {Yee}, {Carlberg}, \&
  {Ellingson}}]{balogh99}
{Balogh}, M.~L., {Morris}, S.~L., {Yee}, H.~K.~C., {Carlberg}, R.~G., \&
  {Ellingson}, E. 1999, \apj, 527, 54

\bibitem[{{Bolatto} {et~al.}(2013){Bolatto}, {Wolfire}, \& {Leroy}}]{bolatto13}
{Bolatto}, A.~D., {Wolfire}, M., \& {Leroy}, A.~K. 2013, \araa, 51, 207

\bibitem[{{Bruzual} \& {Charlot}(2003)}]{bc03}
{Bruzual}, G., \& {Charlot}, S. 2003, \mnras, 344, 1000

\bibitem[{{Cappellari} {et~al.}(2011){Cappellari}, {Emsellem}, {Krajnovi{\'c}},
  {McDermid}, {Scott}, {Verdoes Kleijn}, {Young}, {Alatalo}, {Bacon}, {Blitz},
  {Bois}, {Bournaud}, {Bureau}, {Davies}, {Davis}, {de Zeeuw}, {Duc},
  {Khochfar}, {Kuntschner}, {Lablanche}, {Morganti}, {Naab}, {Oosterloo},
  {Sarzi}, {Serra}, \& {Weijmans}}]{cappellari11}
{Cappellari}, M., {Emsellem}, E., {Krajnovi{\'c}}, D., {et~al.} 2011, \mnras,
  413, 813

\bibitem[{{Cappellari} {et~al.}(2013){Cappellari}, {Scott}, {Alatalo}, {Blitz},
  {Bois}, {Bournaud}, {Bureau}, {Crocker}, {Davies}, {Davis}, {de Zeeuw},
  {Duc}, {Emsellem}, {Khochfar}, {Krajnovi{\'c}}, {Kuntschner}, {McDermid},
  {Morganti}, {Naab}, {Oosterloo}, {Sarzi}, {Serra}, {Weijmans}, \&
  {Young}}]{cappellari13}
{Cappellari}, M., {Scott}, N., {Alatalo}, K., {et~al.} 2013, \mnras, 432, 1709

\bibitem[{{Chabrier}(2003)}]{chabrier03}
{Chabrier}, G. 2003, \pasp, 115, 763

\bibitem[{{Combes} {et~al.}(2007){Combes}, {Young}, \& {Bureau}}]{combes07}
{Combes}, F., {Young}, L.~M., \& {Bureau}, M. 2007, \mnras, 377, 1795

\bibitem[{{Couch} \& {Sharples}(1987)}]{couch87}
{Couch}, W.~J., \& {Sharples}, R.~M. 1987, \mnras, 229, 423

\bibitem[{{Dale} \& {Helou}(2002)}]{dale02}
{Dale}, D.~A., \& {Helou}, G. 2002, \apj, 576, 159

\bibitem[{{Dannerbauer} {et~al.}(2009){Dannerbauer}, {Daddi}, {Riechers},
  {Walter}, {Carilli}, {Dickinson}, {Elbaz}, \& {Morrison}}]{dannerbauer09}
{Dannerbauer}, H., {Daddi}, E., {Riechers}, D.~A., {et~al.} 2009, \apjl, 698,
  L178

\bibitem[{{Davis} {et~al.}(2016){Davis}, {Greene}, {Ma}, {Pandya}, {Blakeslee},
  {McConnell}, \& {Thomas}}]{davis16}
{Davis}, T.~A., {Greene}, J., {Ma}, C.-P., {et~al.} 2016, \mnras, 455, 214

\bibitem[{{Davis} {et~al.}(2013){Davis}, {Alatalo}, {Bureau}, {Cappellari},
  {Scott}, {Young}, {Blitz}, {Crocker}, {Bayet}, {Bois}, {Bournaud}, {Davies},
  {de Zeeuw}, {Duc}, {Emsellem}, {Khochfar}, {Krajnovi{\'c}}, {Kuntschner},
  {Lablanche}, {McDermid}, {Morganti}, {Naab}, {Oosterloo}, {Sarzi}, {Serra},
  \& {Weijmans}}]{davis13}
{Davis}, T.~A., {Alatalo}, K., {Bureau}, M., {et~al.} 2013, \mnras, 429, 534

\bibitem[{{Davis} {et~al.}(2014){Davis}, {Young}, {Crocker}, {Bureau}, {Blitz},
  {Alatalo}, {Emsellem}, {Naab}, {Bayet}, {Bois}, {Bournaud}, {Cappellari},
  {Davies}, {de Zeeuw}, {Duc}, {Khochfar}, {Krajnovi{\'c}}, {Kuntschner},
  {McDermid}, {Morganti}, {Oosterloo}, {Sarzi}, {Scott}, {Serra}, \&
  {Weijmans}}]{davis14}
{Davis}, T.~A., {Young}, L.~M., {Crocker}, A.~F., {et~al.} 2014, \mnras, 444,
  3427

\bibitem[{{Dekel} \& {Birnboim}(2006)}]{dekel06}
{Dekel}, A., \& {Birnboim}, Y. 2006, \mnras, 368, 2

\bibitem[{{Di Matteo} {et~al.}(2005){Di Matteo}, {Springel}, \&
  {Hernquist}}]{dimatteo05}
{Di Matteo}, T., {Springel}, V., \& {Hernquist}, L. 2005, \nat, 433, 604

\bibitem[{{Dressler} \& {Gunn}(1983)}]{dressler83}
{Dressler}, A., \& {Gunn}, J.~E. 1983, \apj, 270, 7

\bibitem[Feldmann et~al.(2017)]{feldmann17} Feldmann, R., Quataert, E., Hopkins, P.~F., Faucher-Gigu{\`e}re, C.-A., \& Kere{\v s}, D.\ 2017, \mnras, 470, 1050.

\bibitem[{{Feldmann} \& {Mayer}(2015)}]{feldmann15}
{Feldmann}, R., \& {Mayer}, L. 2015, \mnras, 446, 1939

\bibitem[{{French} {et~al.}(2015){French}, {Yang}, {Zabludoff}, {Narayanan},
  {Shirley}, {Walter}, {Smith}, \& {Tremonti}}]{french15}
{French}, K.~D., {Yang}, Y., {Zabludoff}, A., {et~al.} 2015, \apj, 801, 1

\bibitem[{{Geach} {et~al.}(2013){Geach}, {Hickox}, {Diamond-Stanic}, {Krips},
  {Moustakas}, {Tremonti}, {Coil}, {Sell}, \& {Rudnick}}]{geach13}
{Geach}, J.~E., {Hickox}, R.~C., {Diamond-Stanic}, A.~M., {et~al.} 2013, \apjl,
  767, L17

\bibitem[{{Genzel} {et~al.}(2015){Genzel}, {Tacconi}, {Lutz}, {Saintonge},
  {Berta}, {Magnelli}, {Combes}, {Garc{\'{\i}}a-Burillo}, {Neri}, {Bolatto},
  {Contini}, {Lilly}, {Boissier}, {Boone}, {Bouch{\'e}}, {Bournaud}, {Burkert},
  {Carollo}, {Colina}, {Cooper}, {Cox}, {Feruglio}, {F{\"o}rster Schreiber},
  {Freundlich}, {Gracia-Carpio}, {Juneau}, {Kovac}, {Lippa}, {Naab}, {Salome},
  {Renzini}, {Sternberg}, {Walter}, {Weiner}, {Weiss}, \& {Wuyts}}]{genzel15}
{Genzel}, R., {Tacconi}, L.~J., {Lutz}, D., {et~al.} 2015, \apj, 800, 20

\bibitem[{{Graves} \& {Schiavon}(2008)}]{graves08}
{Graves}, G.~J., {Schiavon}, R.~P. 2008, \apjs, 177, 446

\bibitem[{{Hopkins} {et~al.}(2006){Hopkins}, {Hernquist}, {Cox}, {Di Matteo},
  {Robertson}, \& {Springel}}]{hopkins06}
{Hopkins}, P.~F., {Hernquist}, L., {Cox}, T.~J., {et~al.} 2006, \apjs, 163, 1

\bibitem[{{Kauffmann} {et~al.}(2003){Kauffmann}, {Heckman}, {White}, {Charlot},
  {Tremonti}, {Peng}, {Seibert}, {Brinkmann}, {Nichol}, {SubbaRao}, \&
  {York}}]{kauffmann03}
{Kauffmann}, G., {Heckman}, T.~M., {White}, S.~D.~M., {et~al.} 2003, \mnras,
  341, 54

\bibitem[{{Kennicutt}(1998a)}]{kennicuttReview}
{Kennicutt}, Jr., R.~C. 1998, \araa, 36, 189

\bibitem[{{Kennicutt}(1998b)}]{kennicutt98}
{Kennicutt}, Jr., R.~C. 1998, \apj, 498, 541

\bibitem[{{Kere{\v s}} {et~al.}(2005){Kere{\v s}}, {Katz}, {Weinberg}, \&
  {Dav{\'e}}}]{keres05}
{Kere{\v s}}, D., {Katz}, N., {Weinberg}, D.~H., \& {Dav{\'e}}, R. 2005,
  \mnras, 363, 2

\bibitem[{{Kriek} \& {Conroy}(2013)}]{kcDust}
{Kriek}, M., \& {Conroy}, C. 2013, \apjl, 775, L16

\bibitem[{{Kriek} {et~al.}(2009){Kriek}, {van Dokkum}, {Labb{\'e}}, {Franx},
  {Illingworth}, {Marchesini}, \& {Quadri}}]{kriek2009}
{Kriek}, M., {van Dokkum}, P.~G., {Labb{\'e}}, I., {et~al.} 2009, \apj, 700,
  221

\bibitem[{{Kriek} {et~al.}(2010){Kriek}, {Labb{\'e}}, {Conroy}, {Whitaker},
  {van Dokkum}, {Brammer}, {Franx}, {Illingworth}, {Marchesini}, {Muzzin},
  {Quadri}, \& {Rudnick}}]{kriek2010}
{Kriek}, M., {Labb{\'e}}, I., {Conroy}, C., {et~al.} 2010, \apjl, 722, L64

\bibitem[{{Le Borgne} {et~al.}(2006){Le Borgne}, {Abraham}, {Daniel},
  {McCarthy}, {Glazebrook}, {Savaglio}, {Crampton}, {Juneau}, {Carlberg},
  {Chen}, {Marzke}, {Roth}, {J{\o}rgensen}, \& {Murowinski}}]{leborgne06}
{Le Borgne}, D., {Abraham}, R., {Daniel}, K., {et~al.} 2006, \apj, 642, 48

\bibitem[{{Leroy} {et~al.}(2015){Leroy}, {Walter}, {Martini}, {Roussel},
  {Sandstrom}, {Ott}, {Weiss}, {Bolatto}, {Schuster}, \&
  {Dessauges-Zavadsky}}]{leroy15}
{Leroy}, A.~K., {Walter}, F., {Martini}, P., {et~al.} 2015, \apj, 814, 83

\bibitem[{{Magdis} {et~al.}(2012){Magdis}, {Daddi}, {B{\'e}thermin}, {Sargent},
  {Elbaz}, {Pannella}, {Dickinson}, {Dannerbauer}, {da Cunha}, {Walter},
  {Rigopoulou}, {Charmandaris}, {Hwang}, \& {Kartaltepe}}]{magdis12}
{Magdis}, G.~E., {Daddi}, E., {B{\'e}thermin}, M., {et~al.} 2012, \apj, 760, 6

\bibitem[{{Mainzer} {et~al.}(2011){Mainzer}, {Bauer}, {Grav}, {Masiero},
  {Cutri}, {Dailey}, {Eisenhardt}, {McMillan}, {Wright}, {Walker}, {Jedicke},
  {Spahr}, {Tholen}, {Alles}, {Beck}, {Brandenburg}, {Conrow}, {Evans},
  {Fowler}, {Jarrett}, {Marsh}, {Masci}, {McCallon}, {Wheelock}, {Wittman},
  {Wyatt}, {DeBaun}, {Elliott}, {Elsbury}, {Gautier}, {Gomillion}, {Leisawitz},
  {Maleszewski}, {Micheli}, \& {Wilkins}}]{mainzer11}
{Mainzer}, A., {Bauer}, J., {Grav}, T., {et~al.} 2011, \apj, 731, 53

\bibitem[{{Mart{\'{\i}}-Vidal} {et~al.}(2014){Mart{\'{\i}}-Vidal}, {Vlemmings},
  {Muller}, \& {Casey}}]{martividal14}
{Mart{\'{\i}}-Vidal}, I., {Vlemmings}, W.~H.~T., {Muller}, S., \& {Casey}, S.
  2014, \aap, 563, A136

\bibitem[{{Martig} {et~al.}(2009){Martig}, {Bournaud}, {Teyssier}, \&
  {Dekel}}]{martig09}
{Martig}, M., {Bournaud}, F., {Teyssier}, R., \& {Dekel}, A. 2009, \apj, 707,
  250

\bibitem[{{Narayanan} {et~al.}(2012){Narayanan}, {Krumholz}, {Ostriker}, \&
  {Hernquist}}]{narayanan12}
{Narayanan}, D., {Krumholz}, M.~R., {Ostriker}, E.~C., \& {Hernquist}, L. 2012,
  \mnras, 421, 3127

\bibitem[{{Saintonge} {et~al.}(2011){Saintonge}, {Kauffmann}, {Kramer},
  {Tacconi}, {Buchbender}, {Catinella}, {Fabello}, {Graci{\'a}-Carpio}, {Wang},
  {Cortese}, {Fu}, {Genzel}, {Giovanelli}, {Guo}, {Haynes}, {Heckman},
  {Krumholz}, {Lemonias}, {Li}, {Moran}, {Rodriguez-Fernandez}, {Schiminovich},
  {Schuster}, \& {Sievers}}]{saintonge11}
{Saintonge}, A., {Kauffmann}, G., {Kramer}, C., {et~al.} 2011, \mnras, 415, 32

\bibitem[{{Sell} {et~al.}(2014){Sell}, {Tremonti}, {Hickox}, {Diamond-Stanic},
  {Moustakas}, {Coil}, {Williams}, {Rudnick}, {Robaina}, {Geach}, {Heinz}, \&
  {Wilcots}}]{sell14}
{Sell}, P.~H., {Tremonti}, C.~A., {Hickox}, R.~C., {et~al.} 2014, \mnras, 441,
  3417

\bibitem[{{Tacconi} {et~al.}(2013){Tacconi}, {Neri}, {Genzel}, {Combes},
  {Bolatto}, {Cooper}, {Wuyts}, {Bournaud}, {Burkert}, {Comerford}, {Cox},
  {Davis}, {F{\"o}rster Schreiber}, {Garc{\'{\i}}a-Burillo}, {Gracia-Carpio},
  {Lutz}, {Naab}, {Newman}, {Omont}, {Saintonge}, {Shapiro Griffin}, {Shapley},
  {Sternberg}, \& {Weiner}}]{tacconi13}
{Tacconi}, L.~J., {Neri}, R., {Genzel}, R., {et~al.} 2013, \apj, 768, 74

\bibitem[{{Wellons} {et~al.}(2015){Wellons}, {Torrey}, {Ma}, {Rodriguez-Gomez},
  {Vogelsberger}, {Kriek}, {van Dokkum}, {Nelson}, {Genel}, {Pillepich},
  {Springel}, {Sijacki}, {Snyder}, {Nelson}, {Sales}, \&
  {Hernquist}}]{wellons15}
{Wellons}, S., {Torrey}, P., {Ma}, C.-P., {et~al.} 2015, \mnras, 449, 361

\bibitem[{{Whitaker} {et~al.}(2012{\natexlab{a}}){Whitaker}, {Kriek}, {van
  Dokkum}, {Bezanson}, {Brammer}, {Franx}, \& {Labb{\'e}}}]{whitaker12}
{Whitaker}, K.~E., {Kriek}, M., {van Dokkum}, P.~G., {et~al.}
  2012{\natexlab{a}}, \apj, 745, 179

\bibitem[{{Whitaker} {et~al.}(2012{\natexlab{b}}){Whitaker}, {van Dokkum},
  {Brammer}, \& {Franx}}]{whitakerMS}
{Whitaker}, K.~E., {van Dokkum}, P.~G., {Brammer}, G., \& {Franx}, M.
  2012{\natexlab{b}}, \apjl, 754, L29

\bibitem[{{York} {et~al.}(2000){York}, {Adelman}, {Anderson}, {Anderson},
  {Annis}, {Bahcall}, {Bakken}, {Barkhouser}, {Bastian}, {Berman}, {Boroski},
  {Bracker}, {Briegel}, {Briggs}, {Brinkmann}, {Brunner}, {Burles}, {Carey},
  {Carr}, {Castander}, {Chen}, {Colestock}, {Connolly}, {Crocker}, {Csabai},
  {Czarapata}, {Davis}, {Doi}, {Dombeck}, {Eisenstein}, {Ellman}, {Elms},
  {Evans}, {Fan}, {Federwitz}, {Fiscelli}, {Friedman}, {Frieman}, {Fukugita},
  {Gillespie}, {Gunn}, {Gurbani}, {de Haas}, {Haldeman}, {Harris}, {Hayes},
  {Heckman}, {Hennessy}, {Hindsley}, {Holm}, {Holmgren}, {Huang}, {Hull},
  {Husby}, {Ichikawa}, {Ichikawa}, {Ivezi{\'c}}, {Kent}, {Kim}, {Kinney},
  {Klaene}, {Kleinman}, {Kleinman}, {Knapp}, {Korienek}, {Kron}, {Kunszt},
  {Lamb}, {Lee}, {Leger}, {Limmongkol}, {Lindenmeyer}, {Long}, {Loomis},
  {Loveday}, {Lucinio}, {Lupton}, {MacKinnon}, {Mannery}, {Mantsch}, {Margon},
  {McGehee}, {McKay}, {Meiksin}, {Merelli}, {Monet}, {Munn}, {Narayanan},
  {Nash}, {Neilsen}, {Neswold}, {Newberg}, {Nichol}, {Nicinski}, {Nonino},
  {Okada}, {Okamura}, {Ostriker}, {Owen}, {Pauls}, {Peoples}, {Peterson},
  {Petravick}, {Pier}, {Pope}, {Pordes}, {Prosapio}, {Rechenmacher}, {Quinn},
  {Richards}, {Richmond}, {Rivetta}, {Rockosi}, {Ruthmansdorfer}, {Sandford},
  {Schlegel}, {Schneider}, {Sekiguchi}, {Sergey}, {Shimasaku}, {Siegmund},
  {Smee}, {Smith}, {Snedden}, {Stone}, {Stoughton}, {Strauss}, {Stubbs},
  {SubbaRao}, {Szalay}, {Szapudi}, {Szokoly}, {Thakar}, {Tremonti}, {Tucker},
  {Uomoto}, {Vanden Berk}, {Vogeley}, {Waddell}, {Wang}, {Watanabe},
  {Weinberg}, {Yanny}, {Yasuda}, \& {SDSS Collaboration}}]{york00}
{York}, D.~G., {Adelman}, J., {Anderson}, Jr., J.~E., {et~al.} 2000, \aj, 120,
  1579

\bibitem[{{Young} {et~al.}(2011){Young}, {Bureau}, {Davis}, {Combes},
  {McDermid}, {Alatalo}, {Blitz}, {Bois}, {Bournaud}, {Cappellari}, {Davies},
  {de Zeeuw}, {Emsellem}, {Khochfar}, {Krajnovi{\'c}}, {Kuntschner},
  {Lablanche}, {Morganti}, {Naab}, {Oosterloo}, {Sarzi}, {Scott}, {Serra}, \&
  {Weijmans}}]{young11}
{Young}, L.~M., {Bureau}, M., {Davis}, T.~A., {et~al.} 2011, \mnras, 414, 940

\bibitem[{{Zabludoff} {et~al.}(1996){Zabludoff}, {Zaritsky}, {Lin}, {Tucker},
  {Hashimoto}, {Shectman}, {Oemler}, \& {Kirshner}}]{zabludoff96}
{Zabludoff}, A.~I., {Zaritsky}, D., {Lin}, H., {et~al.} 1996, \apj, 466, 104

\end{thebibliography}
\end{document}